\documentclass[conference,]{IEEEtran}
\IEEEoverridecommandlockouts

\usepackage{cite}
\usepackage{amsmath,amssymb,amsfonts}
\usepackage{graphicx}
\usepackage{textcomp}
\usepackage{xcolor}
\usepackage[table]{xcolor}
\usepackage{kotex}
\usepackage{adjustbox} 
\usepackage{subcaption}
\usepackage{caption}
\usepackage{booktabs}
\usepackage{tikz,xcolor}
\usepackage{float}
\definecolor{vscodegreen}{RGB}{0,110,0}

\usepackage{algorithm}
\usepackage{circledsteps}
\pgfkeys{
  /csteps/inner color=white,
  /csteps/outer color=black,
  /csteps/fill color=black
}
\pgfkeys{
  /csteps/inner color=white,
  /csteps/outer color=black,
  /csteps/fill color=black
}
\usepackage{setspace}
\usepackage{algpseudocode}

\newcommand{\bnum}[1]{%
  \tikz[baseline=(n.base)]\node (n) [
    circle, fill=black, inner sep=1.2pt
  ] {\textcolor{white}{\scriptsize\bfseries #1}};%
}

\def\BibTeX{{\rm B\kern-.05em{\sc i\kern-.025em b}\kern-.08em
    T\kern-.1667em\lower.7ex\hbox{E}\kern-.125emX}}
\begin{document}

\title{Unified KV Pooling to Accelerate \\Long-Context LLM Serving}

\author{
\IEEEauthorblockN{Minchul Kang\IEEEauthorrefmark{1},
Changyong Shin\IEEEauthorrefmark{1},
Jinwoo Jeong\IEEEauthorrefmark{1},
Jaerim Park\IEEEauthorrefmark{1},
Woohyun Kim\IEEEauthorrefmark{2},\\
Bonyul Gu\IEEEauthorrefmark{2},
Dongwoo Kang\IEEEauthorrefmark{2},
Gyeongsik Yang\IEEEauthorrefmark{3},
Chuck Yoo\IEEEauthorrefmark{1}}
\IEEEauthorblockA{\IEEEauthorrefmark{1}Korea University, South Korea\\
\{mckang, cyshin, jwjeong, jrpark, chuckyoo\}@os.korea.ac.kr}
\IEEEauthorblockA{\IEEEauthorrefmark{2}KT Corporation, South Korea\\
\{woohyun.kim, bonyul.gu, kang.dongwoo\}@kt.com}
\IEEEauthorblockA{\IEEEauthorrefmark{3}Korea University, South Korea\\
g\_yang@korea.ac.kr}
}

\maketitle

\begin{abstract}
Long-context LLM serving requires offloading KV caches to host-memory and SSDs, but existing mechanisms are not designed for such long contexts. We observe significant inefficiencies in current KV caching in long contexts: high serving latency $\sim$30.7 s, exceeding the typical TTFT requirement of 10 s by more than 3$\times$. Our in-depth analysis explains two major reasons: (1) retrieval is serialized through host-memory and SSD, leaving other host-memory modules and SSDs underutilized, and (2) SSD-based KV retrieval spends 84\% of its time in the kernel filesystem rather than actual device access. 
To address the problems, we propose unified KV pooling, which aggregates multiple host-memory modules and SSDs into a single logical pool and distributes KV caches across devices based on their bandwidth. To eliminate the filesystem overhead, we design KV-passthrough, which bypasses the kernel filesystem and directly accesses SSD-resident KV caches from user space via SPDK. 
Across evaluations on LLaMA 3.1-8B, GPT-OSS-20B, and Qwen3-30B-A3B, unified KV pooling reduces TTFT in long-contexts $\sim$4.1$\times$ over state-of-the-art techniques, all making under 10 s. It also reduces blocked I/O time by up to 23.2$\times$ by eliminating filesystem overhead.

\end{abstract}


\section{Introduction}\label{sec:1}

Long-context request serving has become a common scenario in large language model (LLM) services. In LLM serving, the context of a request is the combined sequence of input and output tokens; a request whose context exceeds 16K tokens is generally considered a long-context request~\cite{iclr-retrieval, acl-pam, iclr-longgenbench}. Such requests increasingly appear in real-time applications that process extensive chat histories, retrieved documents, code repositories, or agent trajectories. Examples include conversational agents~\cite{agent1, agent2}, retrieval-augmented search~\cite{rag1}, and code generation services~\cite{code1, code2}. Furthermore, the maximum context lengths of LLMs are expanding rapidly. For instance, LLaMA~3.1~\cite{llama} offers approximately 64$\times$ larger context capacity than its initial release, while GPT~4.5~\cite{gpt} provides 250$\times$ larger context length.

A key mechanism that makes such serving practical is KV caching in the attention layers. During decoding, the serving engine stores previously computed key and value vectors and reuses them for subsequent tokens, thereby avoiding recomputation of past attention states~\cite{attention, vllm}. But the problem is that the KV cache grows linearly with the context length. As a result, the memory footprint becomes large even for a single request. For example, LLaMA-3.1-405B in FP32 requires 0.98~GB of KV cache at 1K context but 123.05~GB at 128K context \cite{intro-3}.

Because GPU memory cannot hold such large KV caches for long contexts, modern serving systems offload them to host-memory or SSD-backed storage. When a decoding step needs an offloaded KV cache, the engine must retrieve it and transfer it back to GPU memory before computation can proceed. As a result, KV retrieval lies on the critical path of decoding latency. In particular, for long-context requests, the KV cache often extends to SSD-backed storage because host-memory alone is insufficient. Therefore, we focus on KV cache offloading on both host-memory and SSDs.

Our motivating experiments (\S\ref{sec:3}) reveal two problems with existing KV cache offloading under long contexts. 
First, serving latency increases significantly with context length: time-to-first-token (TTFT) reaches 30.7 s at 128K context, exceeding the typical requirement of real-world serving systems (10 s \cite{aegaeon}) by more than 3$\times$. 
The primary cause is that existing engines serialize KV retrieval through the host-memory and SSD I/O path, using only a small subset of the available host-memory modules and SSDs while leaving the rest underutilized.
Second, SSD-based KV retrieval suffers from the substantial filesystem overhead. Our profiling shows that 84\% of the total KV retrieval time is spent in filesystem processing, even though the serving engine already knows the target KV objects and does not require general-purpose file operations.

To address both problems, we propose ``unified KV pooling.'' Unified KV pooling aggregates multiple host-memory modules and SSDs into one logical KV pool and distributes offloaded KV caches across pool devices according to each device's bandwidth. This enables parallel KV offloading and retrieval across devices, rather than routing all KV traffic through the serialized storage path. We also design KV-passthrough, which bypasses the kernel filesystem and accesses SSD-resident KV caches directly from user space through SPDK~\cite{spdk}.

The contributions of this paper are as follows.
\begin{itemize}
    \item Show that the current KV cache offloading suffers from two problems in long-context serving: increased serving latency due to the serialized I/O path and large filesystem overhead in SSD-based retrieval.
    \item Introduce unified KV pooling that 1) distributes KV caches across multiple host-memory and SSD devices and 2) removes filesystem overhead through direct user-space SSD access.
    \item Demonstrate $\sim$4.1$\times$ TTFT reduction and $\sim$23.2$\times$ lower blocked I/O time under long-context workloads compared to existing KV cache offloading techniques.
\end{itemize}

\begin{figure}[t]
    \centering
    \includegraphics[width=0.65\linewidth]{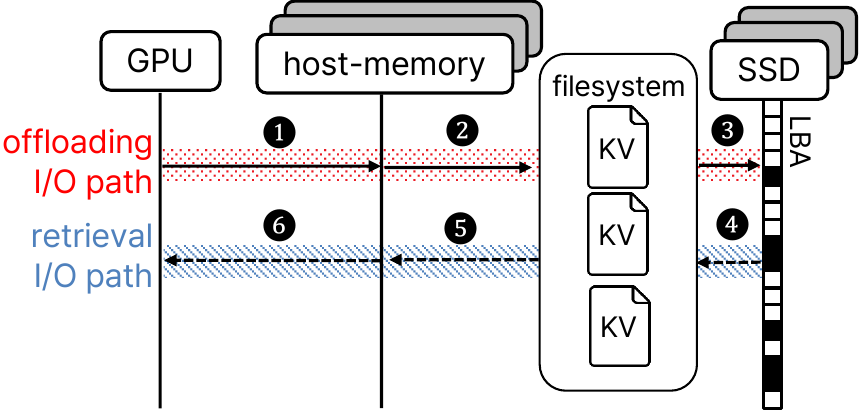}
    \caption{KV cache offloading workflow.}
    \label{fig:bg-1}\vspace{-.5em}
\end{figure}

\section{Background}\label{sec:2}

\subsection{KV Cache Offloading}\label{sec:2.1}

LLM serving engines first reserve GPU memory regions for KV caches. During decoding, each attention layer appends newly generated KV states to the cache and reuses previously stored states for subsequent tokens. As the context length increases, however, the KV cache may exceed the available GPU memory. To avoid out-of-memory failures, modern serving systems offload a portion of the KV cache to SSD-backed storage~\cite{flexgen, lmcache}.

Fig. \ref{fig:bg-1} illustrates the offloading and retrieval workflow.\footnote{We base this explanation on LMCache \cite{lmcache}, a representative engine, when it offloads values on both host-memory and SSD offloading.} During the offloading phase, \bnum{1} evicted KV caches are first transferred from GPU memory to host-memory and \bnum{2} subsequently written to the SSD through the kernel filesystem, \bnum{3} which maps the data to logical block addresses (LBAs) on the storage device. During retrieval, the reverse path is traversed: \bnum{4} the engine reads the corresponding LBAs from the SSD through the filesystem, \bnum{5} stages the data in host-memory, and then \bnum{6} transfers it to GPU memory for decoding. Because the SSD path involves both device I/O and kernel-level filesystem processing, restoring SSD-resident KV caches incurs higher latency than restoring host-memory-resident ones. This directly increases KV retrieval latency, which becomes critical in long contexts where the volume of offloaded KV caches is large.

\begin{figure}[t]
    \centering
    \begin{subfigure}[b]{\linewidth}
        \centering
        \includegraphics[width=0.72\linewidth]{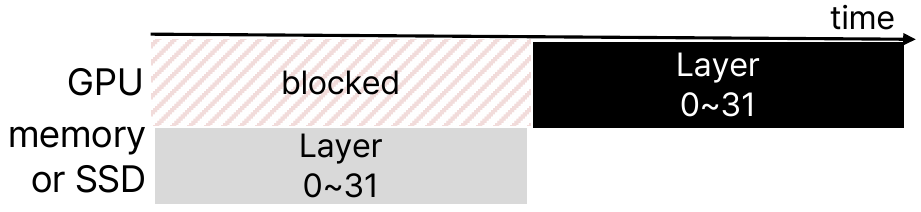}
        \caption{Without layerwise pipelining.}
        \label{fig:bg-2-1}
    \end{subfigure}
    \vspace{0.5em}
    \begin{subfigure}[b]{\linewidth}
        \centering
        \includegraphics[width=0.72\linewidth]{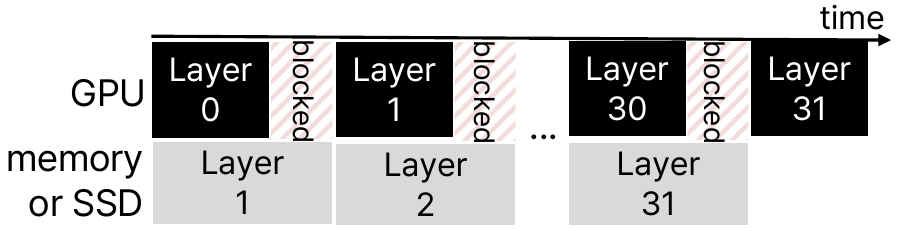}
        \caption{With layerwise pipelining.}
        \label{fig:bg-2-2}\vspace{-.5em}
    \end{subfigure}

    \caption{Layerwise pipelining example. Black box: GPU computation, gray box: KV cache lookup (I/O), Blocked: GPU stall time awaiting KV cache retrieval.}
    \label{fig:bg-2}\vspace{-.5em}
\end{figure}

\subsection{Optimizations in KV Cache Offloading}\label{sec:2.4}

To reduce KV cache retrieval latency, two representative techniques have been proposed: layerwise pipelining and asynchronous KV loading. Recent KV cache management frameworks, including LMCache~\cite{lmcache}, support the techniques.

\subsubsection{Layerwise Pipelining}\label{sec:2.2}

Layerwise pipelining reduces KV retrieval latency by overlapping GPU computation and I/O across layers. Without pipelining, the GPU starts computation of the next layer only after the required KV cache I/O completes (Fig.~\ref{fig:bg-2-1}). With layerwise pipelining, the serving engine computes layer~$N$ while prefetching the KV cache for layer~$N{+}1$, thereby hiding part of the I/O latency (Fig.~\ref{fig:bg-2-2}). 

However, pipelining is effective only when the KV retrieval time is shorter than or comparable to the compute time. Once retrieval takes longer than GPU computation, an uncovered portion appears. We refer to this uncovered portion as blocked I/O time, denoted as \textit{blocked} (Fig. \ref{fig:bg-2-2}): the duration during which only I/O for KV retrieval proceeds, with no overlapping GPU computation. Blocked I/O time directly adds to decoding latency.

\subsubsection{Asynchronous KV Loading}\label{sec:2.3} 
Asynchronous KV loading is a widely used approach~\cite{lmcache, flexgen} to reduce latency by prefetching offloaded KV caches ahead of decoding. In conventional KV caching, KV values are retrieved synchronously in decoding order—each value is fetched only when the corresponding token is about to be processed—making I/O a significant component of decoding time. 
In contrast, asynchronous KV loading issues retrieval requests as soon as a new request arrives and leverages the queuing period before decoding begins. In typical serving systems, incoming requests wait in a queue until they are processed; during this time, the system can identify existing KV caches (e.g., from shared or similar prefixes with previous requests) and prefetch them in advance. So, KV retrieval on some prefix tokens is decoupled from decoding.

\begin{figure*}[t]
    \centering    
    \setbox0=\hbox{\includegraphics[height=0.17\textwidth]{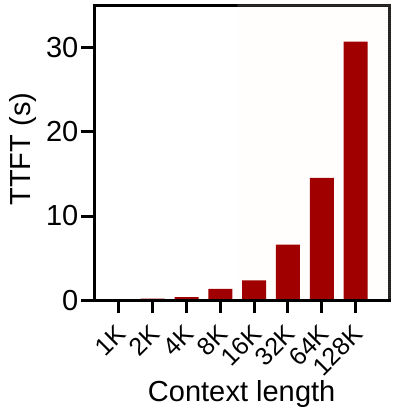}}
    \begin{minipage}[t]{\wd0}
        \centering
        \usebox0\par
        \caption{TTFT.}
        \label{fig:m2}
    \end{minipage}\hfil
    \setbox0=\hbox{\includegraphics[height=0.17\textwidth]{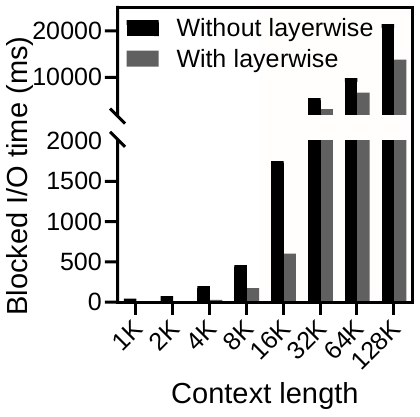}}
    \begin{minipage}[t]{\wd0}        
        \usebox0\par
        \captionsetup{justification=centering, singlelinecheck=false}
        \caption{Blocked I/O\\time.}
        \label{fig:m5}
    \end{minipage}\hfil
    \setbox0=\hbox{\includegraphics[height=0.17\textwidth]{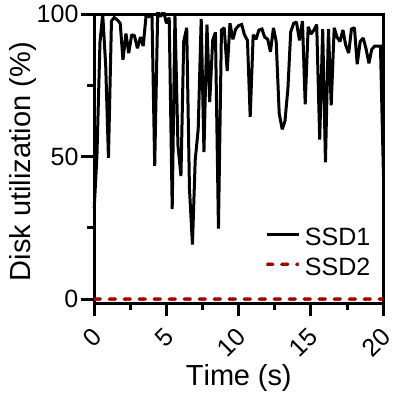}}
    \begin{minipage}[t]{\wd0}        
        \usebox0\par
        \captionsetup{justification=centering, singlelinecheck=false}
        \caption{Unutilized\\SSD.}
        \label{fig:m4}
    \end{minipage}\hfil
    \setbox0=\hbox{\includegraphics[height=0.17\textwidth]{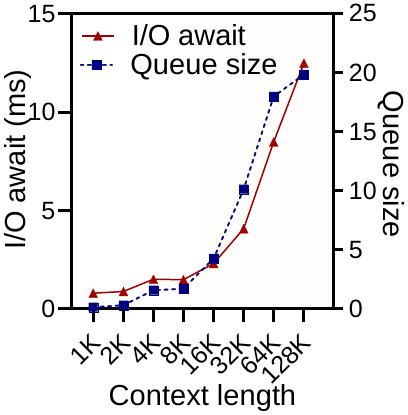}}
    \begin{minipage}[t]{\wd0}
        \centering
        \usebox0\par
        \captionsetup{justification=centering, singlelinecheck=false}
        \caption{Disk\\bottleneck.}
        \label{fig:m3}
    \end{minipage}\hfil
    \setbox0=\hbox{\includegraphics[height=0.17\textwidth]{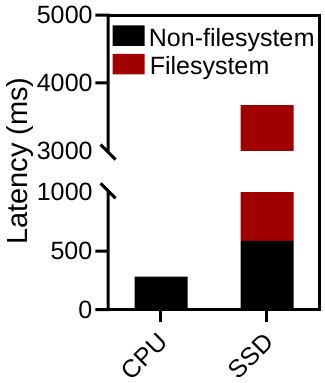}}
    \begin{minipage}[t]{\wd0}
        \centering
        \usebox0\par
        \captionsetup{justification=centering, singlelinecheck=false}
        \caption{Filesystem\\overhead.}
        \label{fig:m6}
    \end{minipage}
    \vspace{-1em}
\end{figure*}

\section{Motivating Experiments}\label{sec:3}
This section demonstrates why the current KV cache offloading becomes inefficient for long-context serving. We identify two problems through motivating experiments: (1) high serving latency stems from a serialized IO path for KV cache (\S\ref{sec:3.2}) and (2) filesystem overhead that dominates SSD-based KV retrieval (\S\ref{sec:3.3}).

\subsection{Setup}\label{sec:3.1}

We use Qwen3-30B-A3B \cite{qwen} on a server equipped with an NVIDIA H100-80GB GPU, two 64 GB DDR4 host-memory modules with a combined peak bandwidth of 21.3 GB/s, and two 2~TB NVMe SSDs connected through PCIe 5.0$\times$16 with a combined peak bandwidth of 63 GB/s.
 
We use vLLM~v0.13.0~\cite{vllm} as serving engine and LMCache~v0.3.10~\cite{lmcache} for KV cache framework. We allocate 20 GB of host memory and 200 GB of SSD storage for offloading and enable LMCache's layerwise pipelining and asynchronous KV loading (\S\ref{sec:2.2}). We evaluate on the Long Doc QA benchmark~\cite{longdoc} and vary the input length from 1K to 128K tokens to cover both short- and long-context requests. We fix the maximum output length to 100 tokens to control variation and enable a fairer comparison across different context lengths. This is consistent with prior work that varies input length under a capped output length \cite{ragcache,flashinfer,solidattention}.

\subsection{Problem~1: High Serving Latency}\label{sec:3.2}

We measure TTFT, the time required to generate the first output token. Fig. \ref{fig:m2} shows TTFT across different context lengths. TTFT increases by $\sim$205.3$\times$ as the context length grows from 1K to 128K, reaching $\sim$30.7 s at 128K---more than 3$\times$ higher than the typical serving requirement of 10 s \cite{aegaeon}. The increase is modest for short contexts but becomes much steeper in long contexts (16K–128K).

To understand the cause, we examine blocked I/O time.\footnote{As GPU computation and I/O are largely overlapped by layerwise pipelining (\S\ref{sec:2.2}), the isolated I/O bottleneck is captured by blocked I/O time.} Fig.~\ref{fig:m5} shows that blocked I/O time grows from 7.1~ms to 13.8~s as the context length increases from 1K to 128K. The I/O path for retrieving offloaded KV caches is therefore the dominant source of the latency increase.
 
We further analyze why I/O time grows so steeply. Fig.~\ref{fig:m4} shows that, on our server with two SSDs, only one SSD is heavily utilized 
while the other remains idle. KV retrieval requests thus accumulate on a single device. Fig.~\ref{fig:m3} confirms the consequence: as context length increases, the average SSD request completion time grows by $\sim$15.4$\times$, while the number of requests waiting in the SSD hardware queue grows by $\sim$132$\times$.
 The root cause is that although the server has multiple host-memory modules and SSDs, existing engines are designed to serialize the requests. Furthermore, the current version of LMCache uses a single set of host-memory module and SSD. In short, the current KV cache offloading has an inherent limitation to exploit retrieval parallelism across devices.

\subsection{Problem 2: Significant Filesystem Overhead}\label{sec:3.3}

Fig.~\ref{fig:m6} shows that reading offloaded KV caches from SSD takes $\sim$11$\times$ longer than reading them from host-memory. To identify the source of this overhead, we profile the SSD retrieval path using \texttt{bpftrace}~\cite{bpftrace}. The result shows that the filesystem processing accounts for 84\% of the total time.
 
This overhead arises because existing KV cache offloading accesses SSD-resident KV caches through the kernel filesystem, which performs general-purpose operations such as file abstraction, metadata handling, and block management. However, offloaded KV caches are internal data objects managed by the serving engine. When the engine retrieves a KV cache entry, it already knows which object to fetch and can identify the stored location from its own metadata. KV retrieval, therefore, does not require the full filesystem path used for general file access. Nevertheless, the current offloading still traverses the entire filesystem stack, adding substantial latency to SSD-based KV retrieval.

In summary, current KV cache offloading is inefficient for long-context serving because it lacks the parallelism across available devices and incurs large filesystem overhead for SSD-resident KV caches.

\begin{figure}[t]
    \centering
    \includegraphics[width=\linewidth]{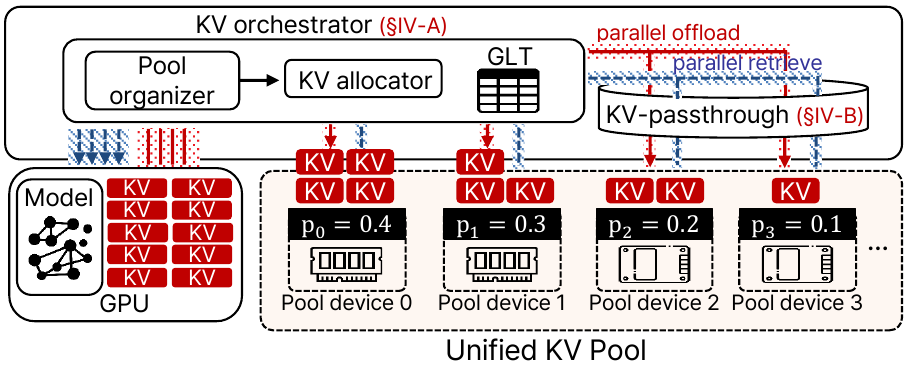}
    \caption{Unified KV pooling structure.}
    \label{fig:design-1}\vspace{-1em}
\end{figure}

\section{Design}\label{sec:4}

To address the two problems identified in \S\ref{sec:3}, we propose unified KV pooling. The design achieves two goals: (1) to exploit device-level parallelism during KV cache offloading and retrieval, and (2) to remove filesystem overhead from SSD-resident KV cache access.
 
Fig.~\ref{fig:design-1} shows the overall architecture. Unified KV pooling aggregates multiple host-memory modules and SSDs into one logical ``KV pool'' for offloaded KV caches. We refer to each host-memory module or SSD in the pool as a \textit{pool device}.
Unified KV pooling has two components: the KV orchestrator (\S\ref{sec:4.1}) and KV-passthrough (\S\ref{sec:4.2}). The KV orchestrator manages KV placement and retrieval across pool devices, ensuring that offloaded KV caches are distributed across multiple devices. KV-passthrough removes the filesystem overhead from SSD-resident KV caches by bypassing the kernel filesystem and issuing direct user-space SSD accesses.

\subsection{KV Orchestrator}\label{sec:4.1}

The KV orchestrator manages offloaded KV caches in the unified KV pool with bandwidth-aware distribution. Instead of concentrating KV traffic on a single device, it assigns KV caches to pool devices in proportion to their bandwidth, preventing any single device from becoming a bottleneck.
It consists of three parts: the global lookup table (GLT), the pool organizer, and the bandwidth-aware KV allocator. The GLT records the placement of each KV cache.
The pool organizer determines the amount of KV traffic each pool device should receive. The bandwidth-aware KV allocator performs offloading and retrieval based on that information.

\subsubsection{Global Lookup Table (GLT)}\label{sec:4.1.1}

The GLT records the location of each offloaded KV cache in the unified KV pool. When the serving engine initializes the model, the KV orchestrator enumerates all pool devices and assigns each one a unique device index. It then initializes the GLT.
 
Each GLT entry stores four fields: (1) a lookup key for the KV cache, (2) the device index of the assigned pool device, (3) the type of the pool device, and (4) the location of the KV cache inside that device. The location is represented as a pointer for host-memory and as a logical block address (LBA) for SSD. Table~\ref{tab:glt} summarizes the fields.

\begin{table}[t]
\centering
\caption{Fields of the global lookup table (GLT).}
\label{tab:glt}\vspace{-.5em}
\begin{tabular}{@{}p{0.2\columnwidth}p{0.76\columnwidth}@{}}
\toprule
\textbf{Field} & \textbf{Description} \\
\midrule
\texttt{key} &
Hash of the KV cache identifier used for lookup \\
\texttt{device\_idx} &
Index of the assigned pool device \\
\texttt{type} &
Pool-device type (\texttt{MEM} or \texttt{SSD}) \\
\texttt{location} &
\parbox[t]{\linewidth}{KV cache location inside the pool device:\\
1) if \texttt{type} is \texttt{MEM}: pointer\\
2) if \texttt{type} is \texttt{SSD}: logical block address (LBA)} \\
\bottomrule
\end{tabular}\vspace{-1em}
\end{table}

\subsubsection{Pool Organizer}\label{sec:4.1.2}

The pool organizer determines how much of the offloaded KV cache should be assigned to each pool device. The goal is to prevent KV traffic from being concentrated on a single device. Instead of using a uniform distribution, the pool organizer distributes KV caches in proportion to device bandwidth, so that faster pool devices receive a larger share of KV I/O.
 
Let $N$ be the number of pool devices, and let $b_i$ be the measured bandwidth of pool device~$i$. The placement ratio $p_i$ of pool device~$i$ is defined as:
\begin{equation}
p_i = \frac{b_i}{\sum_{j=1}^{N} b_j}.
\label{eq:placement_ratio}
\end{equation}
$p_i$ determines the fraction of offloaded KV caches assigned to each pool device. A larger $p_i$ means that the corresponding device receives a larger share of KV traffic. 

\subsubsection{Bandwidth-aware KV Allocator}\label{sec:4.1.3}

The bandwidth-aware KV allocator uses the placement ratios to partition and dispatch offloaded KV caches. At decoding step~$t$, let $\mathcal{K}^{(t)}$ denote the set of KV caches that must be offloaded. The allocator decomposes $\mathcal{K}^{(t)}$ into per-device offload sets and issues writes to the corresponding pool devices in parallel.
 
Let $\mathit{offload\_set}_i^{(t)}$ denote the offload set assigned to pool device~$i$ at step~$t$. Its target size is:
\begin{equation}
|\mathit{offload\_set}_i^{(t)}|
=
\left\lfloor p_i \cdot |\mathcal{K}^{(t)}| \right\rfloor + r_i^{(t)}
\label{eq:target}
\end{equation}
where $r_i^{(t)}$ is the number of remaining KV caches additionally assigned to pool device~$i$ after flooring, subject to:
\[
\sum_{i=1}^{N}
\left(
\left\lfloor p_i \cdot |\mathcal{K}^{(t)}| \right\rfloor + r_i^{(t)}
\right)
=
|\mathcal{K}^{(t)}|.
\]

The allocator first assigns $\lfloor p_i \cdot |\mathcal{K}^{(t)}| \rfloor$ KV caches to each pool device and then distributes the remainder to the devices with larger $p_i$. 
Specifically, the KV caches in $\mathcal{K}^{(t)}$ are enumerated in layerwise order and assigned to offload sets in that same sequence.
Each pool device has its own worker thread, so the resulting writes are issued as independent I/O streams rather than as one serialized path.

\subsection{KV-passthrough}\label{sec:4.2}

KV-passthrough is designed to remove the filesystem overhead from SSD-resident KV cache access. Once an offloaded KV cache is represented in the GLT by its device index and LBA, it no longer needs to be managed through the filesystem as a regular file.
To implement this path, we use SPDK~\cite{spdk}, a user-space NVMe I/O framework. KV-passthrough issues NVMe commands directly from user space and bypasses the kernel filesystem. It exposes two primitives: \texttt{SSD\_Direct\_Store} and \texttt{SSD\_Direct\_Retrieve}.
 
\texttt{SSD\_Direct\_Store} writes an SSD-bound offload set to the target pool device. Once the bandwidth-aware KV allocator assigns an offload set to an SSD pool device, KV-passthrough allocates a contiguous LBA range on the target device and writes the KV caches to that range through asynchronous NVMe write commands. The LBA range length is $\lceil |\mathit{offload\_set}_i^{(t)}| \cdot \text{KV\_size} \;/\; \text{sector\_size} \rceil$ blocks. Writes are issued asynchronously, so command submission is decoupled from completion handling. The resulting LBA information is returned to the bandwidth-aware KV allocator and stored in the GLT.

\texttt{SSD\_Direct\_Retrieve} reads back a KV cache from SSD using the \texttt{device\_idx} and \texttt{location} fields stored in the GLT. From this metadata, KV-passthrough identifies the target LBA range and issues an asynchronous NVMe read. Completion is handled via SPDK polling on a dedicated CPU core, which ensures that KV caching operations are not interfered with by polling overhead. As this path bypasses the filesystem, it removes the overhead identified in \S\ref{sec:3.3}.

\subsection{Workflow Summary}\label{sec:4.3}

\subsubsection{KV Offloading}
 
\begin{algorithm}[t]
\caption{KV cache offloading in unified KV pooling}
\label{algo:offloading}
\begin{algorithmic}[1]
\State \textbf{Input:} offload-required KV set $\mathcal{K}^{(t)}$, pool devices $\mathcal{U}=\{u_1,\dots,u_N\}$, placement ratios $\mathcal{P}=\{p_1,\dots,p_N\}$, GLT
\State \textbf{Output:} updated GLT entries
\Procedure{ParallelOffload}{$\mathcal{K}^{(t)}, \mathcal{U}, \mathcal{P}, GLT$}
\For{$i \gets 1$ \textbf{to} $N$}
    \State $target\_count_i \gets \lfloor p_i \cdot |\mathcal{K}^{(t)}| \rfloor$
    \State $offload\_set_i^{(t)} \gets \emptyset$
\EndFor
\State $remaining\_count \gets |\mathcal{K}^{(t)}| - \sum_{i=1}^{N} target\_count_i$
\For{$j \gets 1$ \textbf{to} $remaining\_count$}
    \State select pool device $u_i$ with the largest $p_i$
    \State $target\_count_i \gets target\_count_i + 1$
\EndFor
\ForAll{$kv \in \mathcal{K}^{(t)}$}
    \State select pool device $u_i$ such that $|offload\_set_i^{(t)}| < target\_count_i$
    \State $offload\_set_i^{(t)} \gets offload\_set_i^{(t)} \cup \{kv\}$
\EndFor
\ForAll{$u_i \in \mathcal{U}$ \textbf{in parallel}}
    \If{$u_i.\texttt{type} = \texttt{MEM}$}
        \State $loc\_map_i \gets \textsc{MemStore}(i, offload\_set_i^{(t)})$
    \Else
        \State $loc\_map_i \gets$
        \Statex \hspace{\algorithmicindent}\hspace{\algorithmicindent}
        $\textsc{SSDDirectStore}(i, offload\_set_i^{(t)})$
    \EndIf
    \State \textsc{UpdateGLT}$(offload\_set_i^{(t)}, i, u_i.type, loc\_map_i)$
\EndFor
\EndProcedure
\end{algorithmic}
\end{algorithm}

Algorithm~\ref{algo:offloading} summarizes the parallel offloading procedure. Lines~4--7 compute the target number of KV caches for each pool device from its placement ratio and initialize the per-device offload sets. Lines~8--12 distribute the remaining KV caches after flooring, prioritizing devices with larger placement ratios. Lines~13--16 partition the KV caches into per-device offload sets. Lines~17--23 issue parallel writes to all pool devices: host-memory pool devices receive writes through direct memory store, while SSD pool devices receive writes through KV-passthrough. After each write completes, the GLT is updated with the device index, device type, and in-device location of each offloaded KV cache.
 
\subsubsection{KV Retrieval}
 
When a decoding step requests offloaded KV caches, the bandwidth-aware KV allocator first looks up their GLT entries to identify the pool device and in-device location of each KV cache. It then groups the requests by pool device and issues parallel reads to the relevant devices.

For host-memory entries, the bandwidth-aware KV allocator retrieves KV caches through direct memory access using the stored pointers. For SSD-resident entries, the allocator invokes \texttt{SSD\_Direct\_Retrieve} to read the corresponding LBAs directly from the SSD without traversing the filesystem. The retrieved KV caches are then returned to the serving engine and transferred to GPU memory for decoding. As the KV caches for each request are grouped and dispatched by pool device, retrieval proceeds in parallel across multiple host-memory modules and SSDs. 

\begin{figure*}[t]
    \centering
    \begin{minipage}[t]{0.48\textwidth}
        \vspace{0pt}
        \centering
        \begin{subfigure}[t]{0.35\linewidth}
            \centering
            {\includegraphics[height=3.4cm]{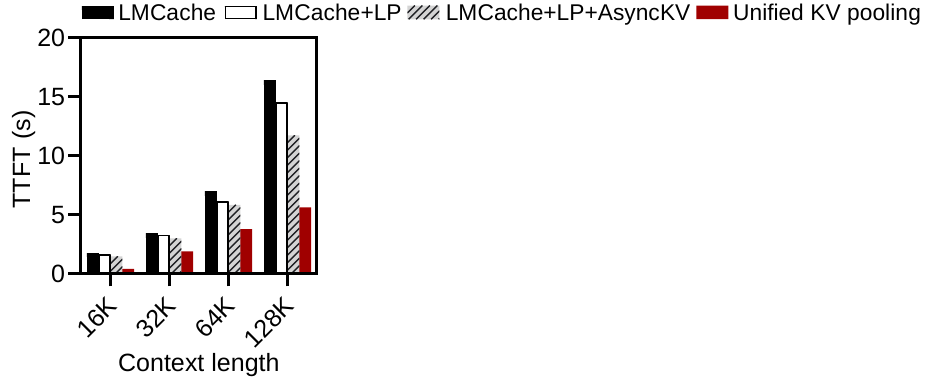}}
            \caption{LLaMA 3.1-8B.}
        \end{subfigure}\hfill
        \begin{subfigure}[t]{0.32\linewidth}
            \centering
            \includegraphics[height=3.2cm]{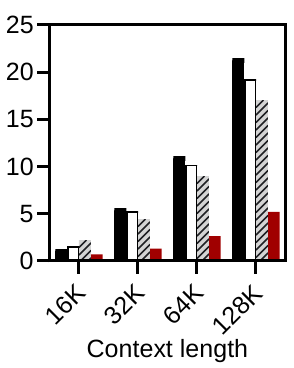}
            \caption{GPT-OSS-20B.}
        \end{subfigure}\hfill
        \begin{subfigure}[t]{0.32\linewidth}
            \centering
            \includegraphics[height=3.2cm]{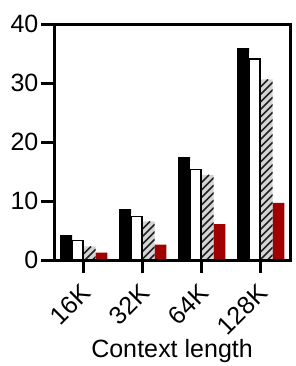}
            \caption{Qwen3-30B-A3B.}
        \end{subfigure}
        \addtocounter{figure}{-1}
        \captionof{figure}{TTFT comparison (\S\ref{sec:5:ttft}).}
        \label{fig:ttft}
    \end{minipage}
    \hfill
    \begin{minipage}[t]{0.48\textwidth}
        \vspace{0pt}
        \centering
        \begin{subfigure}[t]{0.32\linewidth}
            \centering
            {\includegraphics[height=3.45cm]{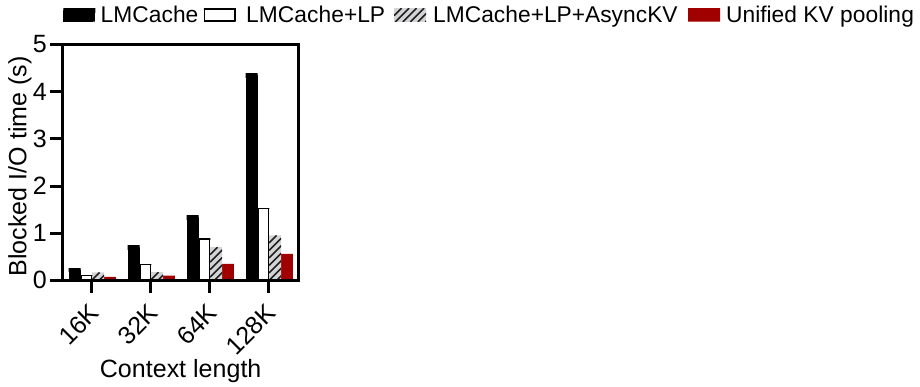}}
            \caption{LLaMA 3.1-8B.}
        \end{subfigure}\hfill
        \begin{subfigure}[t]{0.32\linewidth}
            \centering
            \includegraphics[height=3.2cm]{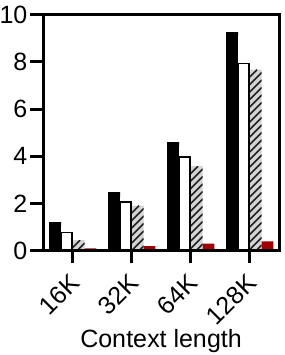}
            \caption{GPT-OSS-20B.}
        \end{subfigure}\hfill
        \begin{subfigure}[t]{0.32\linewidth}
            \centering
            \includegraphics[height=3.25cm]{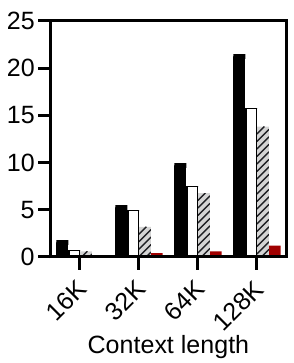}
            \caption{Qwen3-30B-A3B.}
        \end{subfigure}
        \captionof{figure}{I/O overhead (\S\ref{sec:5:blockedio}).}
        \label{fig:io}
    \end{minipage}
    \vspace{-1em}
\end{figure*}

\section{Evaluation}\label{sec:5}

\subsection{Setup}\label{sec:5.1}
We follow a similar experiment setup explained in \S\ref{sec:3.1}. 
We conduct experiments on three different models: LLaMA-3.1-8B \cite{llama}, GPT-OSS-20B \cite{gpt}, and Qwen3-30B-A3B \cite{qwen}. 
We compare three SOTA techniques with unified KV pooling: (1) LMCache, (2) LMCache with layerwise pipelining (LMCache+LP), and (3) LMCache with layerwise pipelining and asynchronous KV cache loading (LMCache+LP+AsyncKV). Layerwise pipelining and asynchronous KV cache loading are recent KV caching optimization techniques (\S\ref{sec:2.4}). We focus on LMCache-based systems as they represent the dominant KV offloading approach in current LLM serving frameworks.

The evaluation consists of three parts.
\begin{itemize}
    \item Main results (\S\ref{sec:5.2}): report TTFT and blocked I/O time, as defined in \S\ref{sec:3}.
    \item Ablation study (\S\ref{sec:5.3}): report TTFT with individual components enabled (e.g., KV orchestrator or KV passthrough) and with the full unified KV pooling design.
    \item Overhead (\S\ref{sec:5.4}): present delay from KV orchestrator
\end{itemize}

\subsection{Main Results}\label{sec:5.2}
\subsubsection{TTFT}\label{sec:5:ttft}
Fig. \ref{fig:ttft} reports TTFT across request lengths. Unified KV pooling consistently achieves the best (lowest) TTFT than all baselines, and the gap widens as the request length increases.
Specifically, on LLaMA 3.1-8B, unified KV pooling achieves a TTFT of 5.6~s at 128K, which is $\sim$2.9$\times$ lower (better) than LMCache, $\sim$2.6$\times$ lower than LMCache+LP, and $\sim$2.1$\times$ lower than LMCache+LP+AsyncKV.
On GPT-OSS-20B, it achieves 5.2~s, corresponding to $\sim$4.1$\times$, $\sim$3.7$\times$, and $\sim$3.3$\times$ reductions over LMCache, LMCache+LP, and LMCache+LP+AsyncKV.
On Qwen3-30B-A3B, it achieves 9.8~s, corresponding to $\sim$3.7$\times$, $\sim$3.5$\times$,and $\sim$3.1$\times$ reductions over LMCache, LMCache+LP, and LMCache+LP+AsyncKV.
The results show that reducing retrieval bottlenecks improves end-to-end serving latency across all context lengths, with larger gains in long-context scenarios.

\subsubsection{Blocked I/O time}\label{sec:5:blockedio}
Fig. \ref{fig:io} presents blocked I/O time across context lengths. 

Unified KV pooling achieves the lowest blocked I/O time among all baselines. 
On LLaMA 3.1-8B, unified KV pooling achieves 0.6~s, which is $\sim$7.8$\times$ lower than LMCache, $\sim$2.7$\times$ lower than LMCache+LP, and $\sim$1.7$\times$ lower than LMCache+LP+AsyncKV.
On GPT-OSS-20B, unified KV pooling achieves $\sim$23.2$\times$ lower blocked I/O time than LMCache, $\sim$19.9$\times$ lower than LMCache+LP, and $\sim$19.2$\times$ lower than LMCache+LP+AsyncKV.
Also, on Qwen3-30B-A3B, unified KV pooling achieves $\sim$17.9$\times$ lower time than LMCache, $\sim$13.2$\times$ lower than LMCache+LP, and $\sim$11.5$\times$ lower than LMCache+LP+AsyncKV.
The results show that unified KV pooling further alleviates the retrieval bottleneck beyond recent KV cache optimization.

\begin{figure}[t]
    \centering
    \begin{minipage}[t]{0.42\columnwidth}
        \centering
        \includegraphics[height=3.3cm, keepaspectratio]{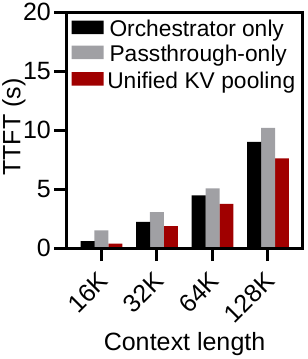}
        \captionof{figure}{Ablation study (\S\ref{sec:5.3}).}
        \label{fig:ablation_1}
    \end{minipage}\hfil
    \begin{minipage}[t]{0.42\columnwidth}
        \centering
        \includegraphics[height=3.3cm, keepaspectratio]{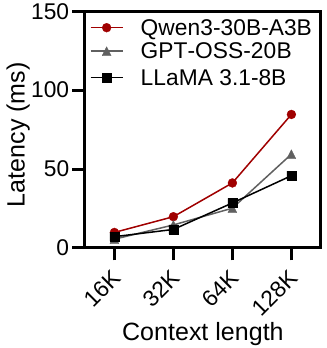}
        \captionof{figure}{Overhead (\S\ref{sec:5.4}).}
        \label{fig:overhead}
    \end{minipage}
    \vspace{-1em}
\end{figure}

\subsection{Ablation Study}\label{sec:5.3}

We evaluate the effect of each component in unified KV pooling by comparing three configurations: (1) KV orchestrator only, which distributes KV caches across pool devices but accesses SSDs through the filesystem; (2) passthrough-only, which bypasses the filesystem but does not distribute KV caches across devices; and (3) the full unified KV pooling, which enables both.
 
Fig. \ref{fig:ablation_1} shows the results. Across all context lengths, the full design outperforms both partial configurations. At 128K, the full design reduces TTFT by $\sim$1.8$\times$ compared to orchestrator only and by $\sim$2.3$\times$ compared to passthrough-only. This result is consistent with the motivation in \S\ref{sec:3}: both KV pooling and KV-passthrough are necessary and removing either significantly degrades performance.

\subsection{Overhead}\label{sec:5.4}

Fig.\ref{fig:overhead} shows the overhead of the KV orchestrator for managing KV placement. Note that this excludes KV cache I/O time and thus reflects only the overhead of unified KV pooling in utilizing memory and storage. As the context length increases, the overhead increases due to increased metadata management and coordination across devices; however, all values remain below 84 ms, which is $\sim$0.8\% of TTFT. 
Note that, despite the overhead reported here, unified KV pooling remains effective for low-latency serving, as demonstrated in \S\ref{sec:5.2} for long-context scenarios, while introducing negligible interference to the overall serving overhead.

\section{Discussion}\label{sec:6}
\noindent\textbf{CPU usage of SPDK.}
Our current implementation relies on SPDK polling for SSD I/O completion. While polling eliminates the kernel interrupt path and reduces I/O latency, it increases CPU utilization due to continuous queue checking. This trade-off is inherent in filesystem-bypassing approaches~\cite{polling-1, polling-2, polling-3}. We dedicate a separate CPU core to polling, so KV caching is not interfered with by polling overhead
This overhead can be mitigated by techniques such as user-level interrupts \cite{spdk+} or hybrid polling. We do not incorporate these techniques, as our focus is on analyzing the KV cache retrieval bottleneck, and leave it to future work.

\noindent\textbf{Multi-turn request serving.}
In this paper, we analyze the limitations of existing techniques and evaluate unified KV pooling on a per-request basis to clearly attribute overheads to KV retrieval from I/O. This setup allows us to exclude interference from request-level concurrency and focus on the fundamental I/O bottleneck in KV cache offloading. In future work, we will extend unified KV pooling to multi-tenant settings where multi-turn requests arrive concurrently.

\section{Related Work}

Various studies have been proposed to improve the performance of KV cache offloading in long-context serving. IMPRESS \cite{impress} mitigates the KV I/O overhead of transferring SSD-resident prefix KVs to the GPU by selectively loading only high-importance KVs identified based on attention weights. InstInfer \cite{instinfer} mitigates the KV I/O bottleneck between SSD and GPU by computing attention directly on a computational storage device, thereby avoiding KV cache transfers to the GPU. InfiniGen \cite{infinigen} reduces the KV cache transfer overhead in long-context serving by estimating the required KV caches for the next layer from the current layer's input and selectively loading them into the GPU. KVQuant \cite{kvquant} reduces the memory and bandwidth burden of long-context inference by quantizing the KV cache to low-bit precision, while CacheGen \cite{cachegen} mitigates KV cache I/O overhead by compressing and transmitting the KV cache.

While these studies primarily focus on reducing the amount of KV cache that must be transferred to the GPU---through selection \cite{impress,infinigen,omnikv}, compression \cite{kvquant,cachegen}, or avoiding GPU transfer \cite{instinfer}---they do not address the bottlenecks targeted in this work, including filesystem overhead in SSD access and limited device-level parallelism in offloaded KV retrieval.
Therefore, they are largely orthogonal to our work.

\section{Conclusion}
This paper proposes unified KV pooling for long-context LLM serving. We show that existing KV caching suffers from severe serving latency degradation in long-context LLM serving due to the serialized I/O path across devices and from filesystem bottlenecks. To address the problems, unified KV pooling aggregates distinct host-memory and SSD devices into a single logical pool and distributes KV caches on the pool, enabling parallel offloading and retrieval. Also, KV-passthrough of unified KV pooling eliminates redundant filesystem overhead. Our evaluation shows that unified KV pooling outperforms existing KV cache offloading techniques, reducing TTFT by $\sim$4.1$\times$ and blocked I/O time by $\sim$23.2$\times$ under long-context workloads.

\bibliographystyle{IEEEtran}
\bibliography{references}

@inproceedings{vllm,
  title={Efficient memory management for large language model serving with pagedattention},
  author={Kwon, Woosuk and Li, Zhuohan and Zhuang, Siyuan and Sheng, Ying and Zheng, Lianmin and Yu, Cody Hao and Gonzalez, Joseph and Zhang, Hao and Stoica, Ion},
  booktitle={Proceedings of the 29th symposium on operating systems principles},
  pages={611--626},
  year={2023}
}

@inproceedings{flexgen,
  title={Flexgen: High-throughput generative inference of large language models with a single gpu},
  author={Sheng, Ying and Zheng, Lianmin and Yuan, Binhang and Li, Zhuohan and Ryabinin, Max and Chen, Beidi and Liang, Percy and R{\'e}, Christopher and Stoica, Ion and Zhang, Ce},
  booktitle={International Conference on Machine Learning},
  pages={31094--31116},
  year={2023},
  organization={PMLR}
}

@article{lmcache,
  title={LMCache: An Efficient KV Cache Layer for Enterprise-Scale LLM Inference},
  author={Cheng, Yihua and Liu, Yuhan and Yao, Jiayi and An, Yuwei and Chen, Xiaokun and Feng, Shaoting and Huang, Yuyang and Shen, Samuel and Du, Kuntai and Jiang, Junchen},
  journal={arXiv preprint arXiv:2510.09665},
  year={2025}
}

@inproceedings{spdk,
  title={SPDK: A development kit to build high performance storage applications},
  author={Yang, Ziye and Harris, James R and Walker, Benjamin and Verkamp, Daniel and Liu, Changpeng and Chang, Cunyin and Cao, Gang and Stern, Jonathan and Verma, Vishal and Paul, Luse E},
  booktitle={2017 IEEE International Conference on Cloud Computing Technology and Science (CloudCom)},
  pages={154--161},
  year={2017},
  organization={IEEE}
}

@article{llama,
  title={The llama 3 herd of models},
  author={Dubey, Abhimanyu and Jauhri, Abhinav and Pandey, Abhinav and Kadian, Abhishek and Al-Dahle, Ahmad and Letman, Aiesha and Mathur, Akhil and Schelten, Alan and Yang, Amy and Fan, Angela and others},
  journal={arXiv e-prints},
  pages={arXiv--2407},
  year={2024}
}

@article{gpt,
  title={gpt-oss-120b \& gpt-oss-20b model card},
  author={Agarwal, Sandhini and Ahmad, Lama and Ai, Jason and Altman, Sam and Applebaum, Andy and Arbus, Edwin and Arora, Rahul K and Bai, Yu and Baker, Bowen and Bao, Haiming and others},
  journal={arXiv preprint arXiv:2508.10925},
  year={2025}
}

@article{qwen,
  title={Qwen3 technical report},
  author={Yang, An and Li, Anfeng and Yang, Baosong and Zhang, Beichen and Hui, Binyuan and Zheng, Bo and Yu, Bowen and Gao, Chang and Huang, Chengen and Lv, Chenxu and others},
  journal={arXiv preprint arXiv:2505.09388},
  year={2025}
}

@inproceedings{impress,
  title={$\{$IMPRESS$\}$: An $\{$Importance-Informed$\}$$\{$Multi-Tier$\}$ Prefix $\{$KV$\}$ Storage System for Large Language Model Inference},
  author={Chen, Weijian and He, Shuibing and Qu, Haoyang and Zhang, Ruidong and Yang, Siling and Chen, Ping and Zheng, Yi and Huai, Baoxing and Chen, Gang},
  booktitle={23rd USENIX Conference on File and Storage Technologies (FAST 25)},
  pages={187--201},
  year={2025}
}

@inproceedings{infinigen,
  title={$\{$InfiniGen$\}$: Efficient generative inference of large language models with dynamic $\{$KV$\}$ cache management},
  author={Lee, Wonbeom and Lee, Jungi and Seo, Junghwan and Sim, Jaewoong},
  booktitle={18th USENIX Symposium on Operating Systems Design and Implementation (OSDI 24)},
  pages={155--172},
  year={2024}
}

@misc{intro-3,
  title        = {Llama 3.1: 405B, 70B \& 8B with multilinguality and long context},
  author       = {{Hugging Face}},
  year         = {2024},
  month        = jul,
  howpublished = {\url{https://huggingface.co/blog/llama31}},
  note         = {Includes an FP16 KV-cache table reporting 0.313\,GB (1K tokens) and 39.06\,GB (128K tokens) for the 70B model. Accessed: 2026-03-02}
}

@misc{longdoc,
  author       = {{LMCache Team}},
  title        = {{LMCache} Benchmarking Documentation: Workload Generator (Long Doc QA)},
  howpublished = {\url{https://docs.lmcache.ai/getting\_started/benchmarking.html}},
  year         = {2026},
  note         = {\textnormal{Accessed: 2026-03-04}}
}

@article{attention,
  title={Attention is all you need},
  author={Vaswani, Ashish and Shazeer, Noam and Parmar, Niki and Uszkoreit, Jakob and Jones, Llion and Gomez, Aidan N and Kaiser, {\L}ukasz and Polosukhin, Illia},
  journal={Advances in neural information processing systems},
  volume={30},
  year={2017}
}

@article{code1,
  title={Evaluating large language models trained on code},
  author={Chen, Mark and Tworek, Jerry and Jun, Heewoo and Yuan, Qiming and Pinto, Henrique Ponde De Oliveira and Kaplan, Jared and Edwards, Harri and Burda, Yuri and Joseph, Nicholas and Brockman, Greg and others},
  journal={arXiv preprint arXiv:2107.03374},
  year={2021}
}

@article{code2,
  title={Code llama: Open foundation models for code},
  author={Roziere, Baptiste and Gehring, Jonas and Gloeckle, Fabian and Sootla, Sten and Gat, Itai and Tan, Xiaoqing Ellen and Adi, Yossi and Liu, Jingyu and Sauvestre, Romain and Remez, Tal and others},
  journal={arXiv preprint arXiv:2308.12950},
  year={2023}
}

@article{rag1,
  title={Retrieval-augmented generation for knowledge-intensive nlp tasks},
  author={Lewis, Patrick and Perez, Ethan and Piktus, Aleksandra and Petroni, Fabio and Karpukhin, Vladimir and Goyal, Naman and K{\"u}ttler, Heinrich and Lewis, Mike and Yih, Wen-tau and Rockt{\"a}schel, Tim and others},
  journal={Advances in neural information processing systems},
  volume={33},
  pages={9459--9474},
  year={2020}
}

@article{agent1,
  title={Hugginggpt: Solving ai tasks with chatgpt and its friends in hugging face},
  author={Shen, Yongliang and Song, Kaitao and Tan, Xu and Li, Dongsheng and Lu, Weiming and Zhuang, Yueting},
  journal={Advances in Neural Information Processing Systems},
  volume={36},
  pages={38154--38180},
  year={2023}
}

@inproceedings{agent2,
  title={Autogen: Enabling next-gen LLM applications via multi-agent conversations},
  author={Wu, Qingyun and Bansal, Gagan and Zhang, Jieyu and Wu, Yiran and Li, Beibin and Zhu, Erkang and Jiang, Li and Zhang, Xiaoyun and Zhang, Shaokun and Liu, Jiale and others},
  booktitle={First conference on language modeling},
  year={2024}
}

@software{bpftrace,
  author       = {{The bpftrace Developers}},
  title        = {bpftrace: High-level tracing language for Linux},
  year         = {2025},
  url          = {https://github.com/bpftrace/bpftrace},
  note         = {Official source repository and documentation. Accessed: 2026-03-09}
}

@inproceedings{spdk+,
  title={SPDK+: Low Latency or High Power Efficiency? We Take Both},
  author={Li, Endian and Yi, Shushu and Peng, Li and Li, Qiao and Zhou, Diyu and Wang, Zhenlin and Wang, Xiaolin and Mao, Bo and Luo, Yingwei and Zhou, Ke and others},
  booktitle={Proceedings of the 17th ACM Workshop on Hot Topics in Storage and File Systems},
  pages={17--23},
  year={2025}
}

@inproceedings{iclr-retrieval,
  title={Retrieval meets long context large language models},
  author={Xu, Peng and Ping, Wei and Wu, Xianchao and McAfee, Lawrence and Zhu, Chen and Liu, Zihan and Subramanian, Sandeep and Bakhturina, Evelina and Shoeybi, Mohammad and Catanzaro, Bryan},
  booktitle={The Twelfth international conference on learning representations},
  year={2023}
}

@inproceedings{acl-pam,
  title={Never lost in the middle: Mastering long-context question answering with position-agnostic decompositional training},
  author={He, Junqing and Pan, Kunhao and Dong, Xiaoqun and Song, Zhuoyang and LiuYiBo, LiuYiBo and Qianguosun, Qianguosun and Liang, Yuxin and Wang, Hao and Zhang, Enming and Zhang, Jiaxing},
  booktitle={Proceedings of the 62nd Annual Meeting of the Association for Computational Linguistics (Volume 1: Long Papers)},
  pages={13628--13642},
  year={2024}
}

@article{iclr-longgenbench,
  title={Longgenbench: Benchmarking long-form generation in long context llms},
  author={Wu, Yuhao and Hee, Ming Shan and Hu, Zhiqing and Lee, Roy Ka-Wei},
  journal={arXiv preprint arXiv:2409.02076},
  year={2024}
}

@article{ragcache,
  title={Ragcache: Efficient knowledge caching for retrieval-augmented generation},
  author={Jin, Chao and Zhang, Zili and Jiang, Xuanlin and Liu, Fangyue and Liu, Shufan and Liu, Xuanzhe and Jin, Xin},
  journal={ACM Transactions on Computer Systems},
  volume={44},
  number={1},
  pages={1--27},
  year={2025},
  publisher={ACM New York, NY}
}

@article{flashinfer,
  title={Flashinfer: Efficient and customizable attention engine for llm inference serving},
  author={Ye, Zihao and Chen, Lequn and Lai, Ruihang and Lin, Wuwei and Zhang, Yineng and Wang, Stephanie and Chen, Tianqi and Kasikci, Baris and Grover, Vinod and Krishnamurthy, Arvind and others},
  journal={Proceedings of Machine Learning and Systems},
  volume={7},
  year={2025}
}

@inproceedings{solidattention,
  title={$\{$SolidAttention$\}$:$\{$Low-Latency$\}$$\{$SSD-based$\}$ Serving on $\{$Memory-Constrained$\}$$\{$PCs$\}$},
  author={Zheng, Xinrui and Wei, Dongliang and Gao, Jianxiang and Song, Yixin and Mi, Zeyu and Chen, Haibo},
  booktitle={24th USENIX Conference on File and Storage Technologies (FAST 26)},
  pages={67--82},
  year={2026}
}

@inproceedings{polling-1,
  title={Faster than flash: An in-depth study of system challenges for emerging ultra-low latency SSDs},
  author={Koh, Sungjoon and Jang, Junhyeok and Lee, Changrim and Kwon, Miryeong and Zhang, Jie and Jung, Myoungsoo},
  booktitle={2019 IEEE International Symposium on Workload Characterization (IISWC)},
  pages={216--227},
  year={2019},
  organization={IEEE}
}

@article{polling-2,
  title={Optimizing storage performance with calibrated interrupts},
  author={Tai, Amy and Smolyar, Igor and Wei, Michael and Tsafrir, Dan},
  journal={ACM Transactions on Storage (TOS)},
  volume={18},
  number={1},
  pages={1--32},
  year={2022},
  publisher={ACM New York, NY}
}

@inproceedings{polling-3,
  title={$\{$UnICom$\}$: A Universally $\{$High-Performant$\}$$\{$I/O$\}$ Completion Mechanism for Modern Computer Systems},
  author={Pan, Riwei and Liang, Yu and Noh, Sam H and Li, Lei and Guan, Nan and Kuo, Tei-Wei and Xue, Chun Jason},
  booktitle={24th USENIX Conference on File and Storage Technologies (FAST 26)},
  pages={721--736},
  year={2026}
}

@inproceedings{omnikv,
  title={Omnikv: Dynamic context selection for efficient long-context llms},
  author={Hao, Jitai and Zhu, Yuke and Wang, Tian and Yu, Jun and Xin, Xin and Zheng, Bo and Ren, Zhaochun and Guo, Sheng},
  booktitle={The Thirteenth International Conference on Learning Representations},
  year={2025}
}

@article{instinfer,
  title={Instinfer: In-storage attention offloading for cost-effective long-context llm inference},
  author={Pan, Xiurui and Li, Endian and Li, Qiao and Liang, Shengwen and Shan, Yizhou and Zhou, Ke and Luo, Yingwei and Wang, Xiaolin and Zhang, Jie},
  journal={arXiv preprint arXiv:2409.04992},
  year={2024}
}

@article{kvquant,
  title={Kvquant: Towards 10 million context length llm inference with kv cache quantization},
  author={Hooper, Coleman and Kim, Sehoon and Mohammadzadeh, Hiva and Mahoney, Michael W and Shao, Yakun S and Keutzer, Kurt and Gholami, Amir},
  journal={Advances in Neural Information Processing Systems},
  volume={37},
  pages={1270--1303},
  year={2024}
}

@inproceedings{cachegen,
  title={Cachegen: Kv cache compression and streaming for fast large language model serving},
  author={Liu, Yuhan and Li, Hanchen and Cheng, Yihua and Ray, Siddhant and Huang, Yuyang and Zhang, Qizheng and Du, Kuntai and Yao, Jiayi and Lu, Shan and Ananthanarayanan, Ganesh and others},
  booktitle={Proceedings of the ACM SIGCOMM 2024 Conference},
  pages={38--56},
  year={2024}
}

@inproceedings{aegaeon,
  title={Aegaeon: Effective GPU pooling for concurrent LLM serving on the market},
  author={Xiang, Yuxing and Li, Xue and Qian, Kun and Yang, Yufan and Zhu, Diwen and Yu, Wenyuan and Zhai, Ennan and Liu, Xuanzhe and Jin, Xin and Zhou, Jingren},
  booktitle={Proceedings of the ACM SIGOPS 31st Symposium on Operating Systems Principles},
  pages={1030--1045},
  year={2025}
}

\end{document}